\documentclass[sigconf,nonacm]{acmart}
 \usepackage{amsmath,amsthm,enumitem}
\usepackage{caption}
\usepackage{multirow}
\usepackage{xcolor}
\usepackage{graphicx}
\usepackage{epstopdf}
\usepackage{url}
\usepackage{comment}
\usepackage{graphicx}
\usepackage{caption}
\usepackage{subcaption}
\usepackage{array}
\usepackage{cleveref}
\newcolumntype{L}{>{\arraybackslash}m{3cm}}
\AtBeginDocument{%
  \providecommand\BibTeX{{%
  \normalfont 
  \kern-0.5em{\scshape i\kern-0.25em b}\kern-0.8em\TeX}
  }}
\begin{document}

\title{Machine Learning-based Relative Valuation of Municipal Bonds}



\author{Preetha Saha}
 \email{preetha.saha@blackrock.com}
 \affiliation{%
   \institution{BlackRock, Inc.}
   \city{New York}
   \state{NY}
   \country{USA}
}

 \author{Jingrao Lyu}
 \email{jingrao.lyu@blackrock.com}
 \affiliation{%
   \institution{BlackRock, Inc.}
   \city{Atlanta}
   \state{GA}
   \country{USA}
}

 \author{Dhruv Desai}
 \email{dhruv.desai1@blackrock.com}
 \affiliation{%
   \institution{BlackRock, Inc.}
  \city{New York}
   \state{NY}
   \country{USA}
}

\author{Rishab Chauhan}
\email{rishab.chauhan1113@gmail.com}
\affiliation{%
  \institution{BlackRock, Inc.}
   \city{Atlanta}
   \state{GA}
   \country{USA}
}

\author{Jerinsh Jeyapaulraj}
 \email{jerinsh244@gmail.com}
 \affiliation{%
   \institution{BlackRock, Inc.}
   \city{New York}
   \state{NY}
   \country{USA}
}

\author{Peter Chu}
 \email{peter.chu@blackrock.com}
 \affiliation{%
   \institution{BlackRock, Inc.}
   \city{New York, NY}
   \country{USA}
 }

 \author{Philip Sommer}
 \email{philip.sommer@blackrock.com}
 \affiliation{%
   \institution{BlackRock, Inc.}
   \city{New York}
   \state{NY}
   \country{USA}
 }

\author{Dhagash Mehta}
 \email{dhagash.mehta@blackrock.com}
 \affiliation{%
   \institution{BlackRock, Inc.}
   \city{New York, NY}
   \country{USA}
}

\renewcommand{\shortauthors}{Saha et al.}
\begin{abstract}
The trading ecosystem of the Municipal (muni) bond is complex and unique. With nearly 2\% of securities from over a million securities outstanding trading daily, determining the value or relative value of a bond among its peers is challenging. Traditionally, relative value calculation has been done using rule-based or heuristics-driven approaches, which may introduce human biases and often fail to account for complex relationships between the bond characteristics. We propose a data-driven model to develop a supervised similarity framework for the muni bond market based on CatBoost algorithm. This algorithm learns from a large-scale dataset to identify bonds that are similar to each other based on their risk profiles. This allows us to evaluate the price of a muni bond relative to a cohort of bonds with a similar risk profile. We propose and deploy a back-testing methodology to compare various benchmarks and the proposed methods and show that the similarity-based method outperforms both rule-based and heuristic-based methods.
\end{abstract}

\maketitle

\graphicspath{ {plots/plots_rf_explainability/} }

\section{Introduction}

Municipal (muni) bonds are debt obligations issued by various state and local government entities \cite{temel2001fundamentals,feldstein2008handbook}. These securities (going forward we use bonds and securities interchangeably) are issued to help finance their operations or fund public operations such as the construction of schools, highways, power plants, bridges, hospitals, etc. The muni market has existed for over 200 years and over the past sixty years it has grown from \$25 billion to its current size of approximately \$4 trillion \cite{sifma2011fundamentals}. The muni market is highly fragmented with over fifty thousand state and local governments issuing debt of various sizes, credit quality, and maturity. While muni bonds are issued in both taxable and tax-exempt formats, the salient quality of being tax-exempt is most attractive for investors.

Less than 10\% of the muni bonds in the universe trade more than 25 days in a year. Moreover, 50\% of the muni bonds trade not more than 5 days in a year. According to the Securities Industry and Financial Markets Association (SIFMA) \cite{sifma2011fundamentals} over the past fifteen years (2008-2022), the average daily number of trades for muni bond market is approximately 39K trades (with approximately an average of 31K retail and 8K institutional trades). The average daily trading dollar value over the same period has been \$13 billion (with approximately an average of \$ 1 billion of retail and \$12 billion of institutional trading dollar value). For 2022, as a percentage of dollar value, institutional trades account for approximately 90\% of daily trades (10\% retail) while only accounting for an average of approximately 18\% of daily trades (82\% retail) \cite{annurev:muni_bond_market,ang2014risk,ang2014muni,schwert2017municipal}. 

In sum, the nature of the muni market poses a set of unique challenges to determine the value and price of municipal bonds for institutional investing\cite{madhavan2022trading,bagley2023pushing}. Equally challenging is identifying the correct set of securities that best satisfy the portfolio's requirement with a small fraction of the market trading on a given day.

Since muni bonds trade sporadically, there are only a handful of observed variables (such as price, yield, etc.). When a bond is put up for sale, assessing the value of that bond is challenging as there may be little observed information for that bond in the past\cite{griffin2023muniprice,schofield2011trading}. One way to overcome this challenge is to use information from similar bonds that have been traded recently. In the present work, we propose a supervised similarity algorithm using CatBoost algorithm \cite{prokhorenkova2018catboost} which uses different attributes of muni bonds and their risk characteristics to learn a distance metric. We then exploit the learned distance metric, or similarities, for the relative-value analysis. This approach goes beyond traditional correlation analysis, offering a more nuanced view of bond characteristics (including their risk profiles) and their interrelationships.

\subsection{Previous Works and Our Contribution}
Relative valuation for companies has been extensively studied \cite{bhojraj2002my,liu2002equity,rhodes2005valuation,bartram2018agnostic,kaustia2015social} (see Ref.~\cite{damodaran2012investment} for a recent review). For mutual funds, supervised similarity was investigated in Refs.~\cite{desai2021robustness, satone2021fund2vec,desai2023quantifying}. For corporate bonds, a supervised similarity approach based on Random Forests was investigated in Ref.~\cite{jeyapaulraj2022supervised}. Ref.~\cite{geertsema2023relative} investigated the relative valuation of companies using Gradient Boosting Decision Trees to eventually value initial public offerings (IPOs). Recently, in Ref.\cite{kolm2022systematic}, a ridge regression and Kalman filter-based approach was proposed to price muni bonds.

Computing similarities between data points can be done in multiple ways. At its core similarity can be defined in terms of the distance between data points, i.e., two or more data points are \textit{similar} if the distance between them is small. Naively, one can start by picking a set of features (in the case of muni bonds, a handful of attributes such as sector, state in which the bond is issued, and time for the bond to mature), and for each data-point sort these features and pick all other data points which satisfy this criteria. Scaling this rule-based approach is difficult: deciding on which attributes to pick first and assigning weights to each of them is not straightforward. 

Alternatively, one can perform unsupervised clustering on the data, though which can help identify patterns in the data, there is no unique objective way to evaluate the quality of clusters nor is assigning weights on individual features \cite{hastie2009elements,bagde2018corpcluster}. 

As opposed to that, a supervised similarity learning algorithm learns the weights of the features given a priori objective of similarity computation (i.e., the target variable). In the present work, we show how tree-based models are powerful methods to learn distance metrics in a supervised manner \cite{jeyapaulraj2022supervised, desai2021robustness, desai2023quantifying}. In the financial industry, supervised similarity learning also has a wide range of applications. It is used for identifying similar companies based on their earnings and attributes, for identifying funds that are similar based on their portfolio composition, and even for corporate bonds \cite{desai2021robustness,jeyapaulraj2022supervised,satone2021fund2vec,vamvourellis2023companysimilarityusinglarge}, etc. using the Random Forest algorithm.

We compare three tree-based algorithms: decision trees, Random Forest, and CatBoost, and show the superiority of CatBoost which is known to perform better on datasets with lots of categorical features. Additionally, we pose the similarity learning problem on multiple targets at once, using multi-output regression. We also propose a novel way to calculate proximity for Boosting trees as compared to existing definitions of Random Forest proximities \cite{rhodes2023geometry}.    

\section{Methodology}
 
In this Section, we describe the proposed methodology. In particular, we start by describing the CatBoost algorithm and then its extension to multi-output regression. We also describe the evaluation metrics as well as explainability method used in the present work.

\subsection{CatBoost Algorithm}
Gradient-boosted decision trees (GBDTs) are a powerful tool for building supervised classification and regression models \cite{natekin2013gradient}. The model constructs a chain of consecutive decision trees, each of which corrects and improves the accuracy based on the errors from its predecessor. Each decision tree learns complex and optimal rule-based decisions from in-sample data without over-fitting, which as a result are very accurate and robust for out-of-sample predictions. The CatBoost \cite{prokhorenkova2018catboost} algorithm is a type of ensemble machine learning method that implements boosting on decision trees, but while the traditional GBDTs is a general framework for ensemble learning, CatBoost is a specific implementation of this framework designed to better handle machine learning tasks involving categorical features and heterogeneous data. 

The CatBoost algorithm starts with an initial constant prediction for all instances (usually 0) and iteratively implements/builds a new decision tree to correct the errors made by the existing model. The iterations can continue until a specified number of trees are added or the error falls below a set threshold. The splits are determined by the feature that maximizes the loss reduction and the threshold for the splits is determined to maximize the cosine similarity between predictions and residual errors. The resultant model is a consecutive sequence of trees where the prediction is a weighted sum of the output from each tree. The key areas where CatBoost is unique from traditional GBDTs are – (1) the model treats the data sequentially while doing target encoding and while calculating output values from the trees to avoid data leakage; (2) it creates large symmetric trees which are weaker learners with fast computation times; (3) internally implements and encodes feature combinations of categorical variables as input variables such that in a current tree, the model combines all categorical features already used in the previous split with all categorical features in the dataset.

\subsection{Similarity learning using CatBoost Model }
In the context of similarity or proximity learning in the tree-based models, each leaf node of a decision tree signifies a division in the feature space. All the instances within the same leaf node can be considered as the nearest neighbors of any other data point within the same division \cite{breiman-cutler-blog,rhodes2023geometry}. The predicted value of the target variable for a data point within this division is computed as the average of the target values of all the training data present in the same division. For the RFs and GBDTs, this prediction is further aggregated across all trees and multiplied by their weights. This approach allows the model to capture the inherent structure and relationships within the data that can be used to compute local distances between instances as opposed to only global distances. For the RFs, proximity between pairs of instances is proportional to the number of times the instances fall into the same leaf node across all trees. The similarity score between instances $i$ and $j$ is calculated using Eq.~(\ref{eq:rf_proxomity}):
\begin{equation}
 S(i,j)  = \frac{1}{N}\sum_{t=1}^{N} I_t(v(j)=v(i))
\label{eq:rf_proxomity}
\end{equation}
where  $N$ is the number of trees in the trained RF, $I$ is the indicator function if the leaf index $v(i)$ of instance $i$ is same as that of instance $j$ in tree $t$.

For the GBDTs, each tree learns to predict the residuals from its predecessor, hence the trees earlier in the sequence have a larger impact in correcting the overall amplitude of errors in predictions than the trees later in the sequence. The sequence of trees has an approximately exponentially decreasing importance in the context of error correction for any given prediction. The magnitude of the difference in residual error correction from one tree to the next is used as a proxy to quantify tree importance. We compute the similarity score between pairs of instances using the CatBoost algorithm as follows,
\begin{equation}
 P(i,j)  = \frac{1}{N}\sum_{t=1}^{N} \Delta_t I_t(v(j)=v(i))
\label{eq:gbm_proxomity}
\end{equation}
where  $\Delta_t$ is \textit{tree importance} proportional to the difference in residual errors between consecutive trees, i.e.,
\begin{equation}
\Delta_t =  \frac{ E(t) - E(t-1) }{\sum_{t=1}^{N}  E(t) -  E(t-1) },
\label{eq:tree_imp}
\end{equation}
where $E(t)$ is the error from model after adding tree $t$. We use the MultiRMSE as our error metric.

\subsubsection{Multi-output Regression}
In a muti-output regression, a model is trained to predict multiple dependent variables simultaneously from a given set of input features that minimizes loss across multiple targets at once, instead of building a separate model for each target. This helps the model leverage the correlations between these different target variables to improve predictions. CatBoost supports multi-output regression with two built-in objective functions for optimization using the Multiple Target Root Mean Squared Error (Multi-RMSE) across multiple dimensions or targets as calculated by the following equation:
\begin{equation}
 MultiRMSE  = \sqrt{\frac{\sum_{i=1}^{N}\sum_{d=1}^{D}(P_{i,d} - y_{i, d})^{2} w_{i}}{\sum_{i=1}^{N} w_{i}}},
 \label{eq: multirmse}
\end{equation}
where $N$ is the total number of instances, $D$ is the number of target variables or the dimension of the label, $P_{i,d}$ is the predicted value, $y_{i,d}$ is the actual value, and $w_{i}$ is the sample weight of instance $i$.

\subsection{Evaluation Metrics for the Regression Phase}
In the present work, we will have a regression task at hand as our target variables will be Options-Adjusted Spread (OAS) and yield which are numerical variables. For the regression task, we used the standard evaluation metrics \cite{hastie2009elements} such as mean squared Error(MSE), mean absolute error(MAE), mean absolute percentage error(MAPE) and $R^2$ to measure the performance of the individual models. If $y_i$ and $\hat{y}_i$ are the observed and predicted values of the target variable for $i$-th data-point, the errors are given by the following equations.

\subsection{Feature Importance using SHAP}
We used SHapley Additive exPlanations (SHAP) to extract the feature importance from the trained model. SHAP assigns each feature in a model an importance value based on its contribution to the model’s prediction \cite{molnar2020interpretable}. It uses Shapley value - a concept from cooperative game theory that measures the marginal contribution of each feature considering all possible combinations of features. We use recent implementations of SHAP for tree-based methods \cite{lundberg2017shap,yang2021fast} for this work.

\subsection{Relative Valuation Methodology}
We first define broad groups of bonds based on two attributes, the state in which the bond is issued and the maturity bucket in which the bond falls (we define ten buckets from zero to ten years to maturity of a bond with a yearly increment and two additional buckets for ten-to-fifteen years and fifteen-and-above years). We call each of these broad groupings a \textit{generic group}\footnote{Due to the limited inventory of available muni bonds at any given point, identifying orders based on characteristics is easier as compared to looking for exact bonds.} (for example, California's ten-year generic group represents all bonds issued by the state of California which mature between nine to ten years from the day of calculation). 

One can argue that we can add more bond characteristics as our filtering criteria and keep narrowing our search space. By doing so we would miss out on finding better investment opportunities on an overall basis. The choice of using state and maturity gives us a constrained yet broad starting point. All the bonds satisfying the generic grouping constraints are also called \textit{perfect matches}. We perform relative-value analysis for each of these generic groups using three different approaches. We validate our results using a back-test over 4 years. Next, we describe our ranking approach and back-testing mechanism.

\subsection{Ranking Approaches}
To initialize our ranking approach we start by looking at all available inventory for each generic group on a given day. We isolate available inventory to its generic group and rank each bond within its generic group using the approaches listed below. Our back-test aims to validate if the top-ranked bonds by the different approaches also produce top returns in these generic groups and if so, how well? 

\subsubsection{Ranking by Yield:}\label{sec:rank_yield} 
The ranking by yield approach simply ranks each of the muni bonds in a given generic group based on the quoted yield of the bond (ranked from highest to lowest yield). This ranking, though straightforward and very efficient from the computational point of view, does not take into account any risk adjustment taken on by trading the highest-yield bond. Hence, this method is not directly applicable to the purpose of the present problem, however, we use this as a very simplistic baseline method keeping in mind that this option may come out to be a better one in the short term and in favorable market conditions.

\subsubsection{Ranking by Duration-times-spread (DxS) (Rule-based Cohort):}\label{sec:rank_dxs} 
The rule-based cohort method identifies bonds with similar risk profiles, as measured by their DxS values. Here, for each bond in a given generic group, we pick $k$ nearest neighbors (i.e., bonds with the closest DxS values to the given bond). Next, for each bond within the generic group, we compute the difference (called the relative value of the bond) of the yield of the bond to the median yield of its $k$ nearest neighbors chosen based on DxS. Finally, we rank all the bonds within the generic group based on the relative value in descending order. Ranking logic relies on the assumption that the largest relative value signifies that the bond is farthest from the cluster in terms of yield (most undervalued) and should provide higher returns over time.

\subsubsection{Ranking by CatBoost Similarity:}\label{sec:rank_sim} Here, we identify cohorts of top $k$ nearest neighbors for each bond within a generic group using the similarity matrix computed using Eq.~(\ref{eq:gbm_proxomity}) with the help of a trained CatBoost model. In this work, we argue that such cohorts are more representative of the target bonds in terms of similar risk profiles and attributes. Similar to the DxS based ranking method, for each generic group, we pick $k$ nearest neighbors using the similarity scores from the trained CatBoost model. Then, similar to the previous section, we rank the bonds in the generic group based on the difference between their yield and the median yield of its $k$ neighbors again in descending order. In the present work, we use $k \in \{ 5, 10, 50, 100 \}$.

\subsection{Back-testing method}
Our objective for the back-test is to validate the different relative-valuation methods, i.e., ranking approaches described in Sections \ref{sec:rank_yield}, \ref{sec:rank_dxs} and \ref{sec:rank_sim}). That is, to check if a bond that is priced cheaper compared to its \textit{true} neighbors tends to perform better or at par compared to its cohort, and if so by how much?

Our testing methodology involves evaluating and comparing the performance of the ranking methods over historical data to assess their effectiveness in identifying undervalued securities. To analyze if relative-value-based ranking contributes to higher returns over time, we run our back-test for a continuous stretch of six months spread independently over five years: We begin our back-test at time period $\mathbf{T}$ and re-evaluate our objective at time periods $\mathbf{T}+1$, $\mathbf{T}+2$, $\mathbf{T}+3$ and $\mathbf{T}+6$ months. Next, we describe how ranking is done at each time period.

\subsubsection{Initial ranking:} For each independent six-month time period, we start by looking at all trade orders daily for the first month and map them to individual unique generic groups. Every bond captured here acts as our test set, we do this over a month to capture different orders to avoid bias for a specific day. Then, for each generic group, we rank the set of matched bonds daily using the three ranking approaches mentioned in Sections \ref{sec:rank_yield}, \ref{sec:rank_dxs} and \ref{sec:rank_sim}. We refer to these ranks as our initial ranks at date $\mathbf{T}$.

\subsubsection{Final ranking:} For each generic group on a given date $\mathbf{T}$, we assign ranks to all bonds in our test set at periods $\mathbf{T}+1$, $\mathbf{T}+2$, $\mathbf{T}+3$ and $\mathbf{T}+6$. These ranks are based on their yield changes from time $T$. We use yield changes as our proxy to measure returns. The objective here is to validate the hypothesis that bonds ranked higher based on the three ranking approaches (i.e., far from their actual cohort at $\mathbf{T}$) tend to perform better on an average over time. This is because we expect these bonds to self-correct in an efficient market.

\subsection{Evaluating Back-test}
To evaluate the effectiveness of different ranking approaches, for each testing period, we calculate two key percentages after periods $\mathbf{T}+1$, $\mathbf{T}+2$, $\mathbf{T}+3$ and $\mathbf{T}+6$ months. The final data is aggregated over generic grouping orders for the first month of each test period. For each time period the two percentages are calculated as follows:
\begin{enumerate}
    \item The percentage of top-ranked bonds for each generic group is in the top three relative returns for that generic group. This metric represents the proportion of generic group for which the first-ranked bond (sorted based on one of the three different ranking methods), is also among the top three highest-ranked bonds in that generic group after a given period. For instance, a value of 7.77\% in the $\mathbf{T}+1$ months indicates that 7.77\% of total generic groups had their first-ranked match among the top three performers in terms of returns after one month.

    \item Percentage of top returning bond for each generic-grouping originated from top three rankings of that method. This metric reflects the proportion of generic groups where the highest returning bond at a given period was present in the top three ranks in the initial selection for that ranking method. For example, a value of 6.28\% in the $\mathbf{T}+1$ months means that 6.28\% of generic groups had their highest returning bond after one month among the top three initial ranks for that ranking method.
\end{enumerate}

For our final evaluation, we average these two percentages to derive a combined back-testing metric. This approach provides a comprehensive assessment of the ranking approaches. The box plots in Figure \ref{fig:alltests} illustrate the distribution of these percentages overall generic groups for each testing month. We perform this calculation for four different values of nearest neighbors $k$ as noted in the previous section. This combined metric helps us address the question: Do the top rankers by different approaches also produce top returns, and if so, to what extent?

\section{Data Description}
Our universe of muni bonds is a subset of the muni Bond market ($\sim225K $ securities), for each snapshot under consideration, as a result of a series of filters around credit quality, small deal sizes, and coupons to exclude high-risk securities. For training the CatBoost algorithm, we use cross-sectional static information on the securities starting from Oct 25th till Nov 1st, 2023. We use OAS computed using an internal proprietary model and yields from BVAL as target variables to train the model.
\begin{table}[htbp]
    \centering
    \begin{tabular}{l l p{1.3in}}
    \toprule
    Feature                 & DataType          & Description\\ \midrule
    State                   & Categorical       & Issuing State \\
    Days-to-maturity        & Numerical         & Remaining days to maturity \\
    Age                     & Numerical         & Days since issue \\
    Coupon                  & Numerical         & \% Annual interest paid \\  
    Coupon Frequency        & Numerical         & Times coupon is paid in a year \\
    Bonds by obligor        & Numerical         & Total active bonds issued by the obligor \\
    Amount Issued           & Numerical         & Issue size of offering \\
    Rating                  & Categorical       & Composite rating for bond \\
    Time-to-call            & Numerical         & Number of days to next call date \\
    Tax Status              & Categorical       & Tax status of offering \\
    Sector Code             & Categorical       & Municipal bond sector \\
    Put-Call                & Categorical       & Optionality of bond \\
    Funding                 & Categorical       & Funding type \\
    Deal Amount             & Numerical         & Original deal amount for the issue \\
    Use of proceeds         & Categorical       & Use of proceeds from bond sale \\
    Payment Frequency       & Categorical       & Frequency at which payments will be made \\
    \bottomrule 
    \end{tabular}
    \caption{Feature set description}
    \label{tab:bond_features}
    \vspace{-6mm}
\end{table}

\subsubsection{Data Pre-processing}
We winsorize the data for Yield and OAS to remove outlier samples. The transformed dataset consisted of a total of 22 features including 11 categorical features and 11 numerical features. We let the CatBoost implementation encode the categorical variables.

\subsection{Trading Data and Back-test period}
For the back-testing analysis, we utilized BVAL pricing and internal trading data of muni bonds for the following periods: June 2019 - December 2019, March 2020 - September 2020, March 2021 - September 2021, July 2022 - January 2023, March 2023 - September 2023 and November 2023 - February 2024. By choosing these periods, we aim to evaluate our approach in different market conditions. Tables \ref{tab:generic_groups_year} and \ref{tab:bonds_per_generic_group} provide basic statistics of our sampled data for our experiments from a larger universe.

Muni market has provided good returns compared to the treasury from 2012 until 2021. From 2019, muni market has rallied (yields have gone down) with lower interest rate environments until 2021. Then, as the US treasury rate sharply increased in 2022, muni yields had a sharp increase and the spread between treasury and muni has diminished. Recently, muni market has further rallied to drop the yield lower in late 2023 and early 2024. The back-testing analysis covers five distinct scenarios that encompass a spectrum of market conditions, including the post and pre-pandemic regimes as well as environments characterized by low, escalating, and high interest rates.  
 
\begin{table}[ht]
    \centering
    \begin{tabular}{l l l}
    \toprule
    Year & \# Generic Groups & Median Bonds per generic group \\ \midrule
    2019 & 649  & 233 \\
    2020 & 486  & 75 \\
    2021 & 649  & 105 \\
    2022 & 839  & 151 \\
    2023 & 1118 & 195 \\
    2024 & 1279 & 248\\ \bottomrule
    \end{tabular}
    \caption{Breakdown for average generic groups per year and median bonds per generic group}
    \label{tab:generic_groups_year}
    \vspace{-6mm}
\end{table}

\begin{table}
    \centering
    \begin{tabular}{ccccc}
    \toprule
         &  Average&  Median&  Min& Max\\
    \midrule
         2019&  233.7&  65.5&  1& 1550\\
         2020&  213&  59.5&  1& 1577\\
         2021&  150.2&  41.5&  1& 1611\\
         2022&  188.2&  43.5&  1& 1649\\
         2023&  164.1&  35&  1& 1648\\
         2024&  364.1&  64.5&  1& 4112\\
    \bottomrule
    \end{tabular}
    \caption{Number of bonds per generic group.}
    \label{tab:bonds_per_generic_group}
\end{table}

\section{Computational Details}
We used CatBoost algorithm as implemented in the publicly available official implementation \url{https://catboost.ai/}. When benchmarking the performance of the CatBoost algorithm with respect to baseline models such as elastic net regression, Decision Tree and Random Forest, we used their single output regression versions as implemented in Scikit-learn \cite{scikit-learn}. For all the evaluations and analysis, we used Pandas library \cite{mckinney2011pandas} in Python.

Our model has undergone training including hyperparameter optimization using cross-validation and sample weighting to ensure robustness and accuracy. We implemented sample weighting allowing the model to pay more attention to certain samples during training. We utilized six months of MSRB trades\cite{msrb2012municipal} data leading upto Nov 1st (our training data period) to weigh our training samples. In this specific scenario, we assigned linear weights to the securities based on each of their latest traded day before Nov 1st. Such that the securities that traded closer to Nov 1st had higher weighting than the ones traded a while ago. This approach allowed the model to give more importance to recent data, thereby enhancing its predictive performance and relevance to current trends.

We evaluated and benchmarked the model performance using various error metrics and against different models. The underlying sections describe the modeling steps in detail. To ensure robustness in our model, after randomized shuffling of the dataset, we selected 80\% of the data for training and the remaining 20\% data for testing. We used randomly chosen 20\% of the training data as a validation set and for hyperparameter optimization.

For CatBoost, We performed hyperparameter tuning for the number of estimators (trees), maximum depth of each tree, maximum features used to decide optimum splits (quality of split calculated using squared error), minimum samples at every leaf node, and learning rate. The range of hyperparameters search space is shown in Table \ref{tab:hyperparameters}. We also tuned the Random State parameter to prevent the model from getting stuck in local minima. We used similar hyperparameter optimization for elastic nets (two hyperparameters $l_1$ and $l_2$), Decision Trees (maximum depth and minimum samples in each leaf node), and RFs (number of estimators, maximum depth, number of leaves in each leaf node and number of random features for each tree). We used Optuna \cite{akiba2019optuna} to tune the hyperparameters. It implements Bayesian optimization with an algorithm called Tree-structured Parzen Estimator.
\begin{table}[htbp]
    \centering
    \begin{tabular}{l l }
    \toprule
    Hyper-Parameter & Search Space \\ \midrule
    \# Estimators   & Numerical [100-2000, step size 25]  \\
    Max Depth       & Numerical [5-25] \\
    \# Leaves & Numerical [8-128] \\
    Max Features    & Numerical [40 - 80\% of total features] \\
    Min Samples     & Numerical [5-100, step size of 25] \\
    Learning Rate   & Numerical [0.001 - 0.2] \\
    Random State    & Numerical [Random integer] \\  \bottomrule
    \end{tabular}
    \caption{Hyper-Parameter search space for CatBoost model}
    \label{tab:hyperparameters}
    \vspace{-6mm}
\end{table}

\section{Results}
In this Section, we describe the results of our experiments.
\subsection{Performance of CatBoost Regressions}
We trained the CatBoost algorithm with Multi-RMSE error metric as in Eq.(~\ref{eq: multirmse}) for the two target variables, OAS and yield, for each security, and trained elastic net regression, Decision Trees and RF, on one target variable at a time. The resulting errors from the model in the train, validation, and test datasets are shown in Table \ref{tab:rmse_all_models}.
The results show that overall CatBoost significantly outperforms all other models.
\begin{table*}[htbp]
    \centering
    \begin{subtable}{0.5\textwidth}
        \centering
        \begin{tabular}{l l l l l l}
        \toprule
          & Fold &  R2&MAE & MSE& MAPE \\ 

        \midrule
        Elastic Net & Train &  0.61 &21.72 &38.56  &32.74 \\
                          &Valid &  0.61 &21.78 &38.79 &53.05 \\
                          &Test & 0.61 &37.26 &39.40 & 1.9 \\
        \midrule

        Decision Tree & Train &  0.90 &9.52 &18.98  &7.53 \\
                          &Valid &  0.755 &15.24 &30.83 &11.11 \\
                          &Test & 0.765 &15.20 &30.59 &1.11 \\
        
        \midrule
        Random Forest & Train &  0.899 &9.36 &19.95  &16.27 \\
                          &Valid &  0.831 &12.78 &25.63 &39.79 \\
                          &Test & 0.833 &12.88 &25.74 &0.86 \\
        \midrule

        CatBoost & Train   &  0.90 & 10.29  & 19.21  & 20.06 \\
                & Valid   &  0.87 & 12.17  & 22.11  & 74.97 \\
                & Test    &  \textbf{0.87} & \textbf{12.26}  & \textbf{22.54}  & \textbf{0.88} \\
                    
        \bottomrule
        \end{tabular}
        \caption{OAS as the target variable.}
        \label{tab:rmse_all_models_oas}
    \end{subtable}%
    \hfill
    \begin{subtable}{0.5\textwidth}
        \centering
        \begin{tabular}{l l l l l l}
        \toprule
             & Fold &  R2&MAE & MSE& MAPE \\ 
            \midrule

            Elastic Net & Train &  0.77 &0.23 & 0.37  &0.051 \\
                              &Valid &  0.76 &0.241 & 0.37 &0.051 \\
                              &Test & 0.76 & 0.242 & 0.38 & 0.051 \\
            \midrule

            Decision Tree & Train &  0.94 & 0.094 & 0.18  &0.02 \\
                              &Valid &  0.86 & 0.15 & 0.29 &0.031 \\
                              &Test & 0.87 & 0.15 & 0.282 &0.031 \\
        
            \midrule   
            Random Forest & Train &  0.93 & 0.098 & 0.20  &0.020 \\
                              &Valid &  0.89 & 0.129 & 0.251 &0.0287\\
                              &Test & 0.897 & 0.13 & 0.252 & 0.027 \\

            \midrule
            CatBoost & Train &  0.94 &0.109 &0.18  &0.023 \\
                     & Valid &  0.93 &0.122 &0.207 &0.026 \\
                     & Test & \textbf{0.93} &\textbf{0.122} & \textbf{0.205} & \textbf{0.025} \\
            \bottomrule
        \end{tabular}
        \caption{Yield as the target variable.}
        \label{tab:rmse_all_models_yield}
    \end{subtable}
\caption{Performance of CatBoost regression for the given data for the snapshot of 1st Nov 2023, for baseline models such as elastic net regression, Decision Tree, and Random Forest. Bold text denotes the best-performing model for the test dataset, showing that overall CatBoost outperforms all other baseline algorithms. Other snapshots exhibit similar results.}
\label{tab:rmse_all_models}
\vspace{-6mm}
\end{table*}

Table \ref{tab:test_evalmetric} summarizes the performance of CatBoost over various snapshots where the model is trained from scratch for each snapshot on the training data within the snapshot and tested on the holdout dataset from the same snapshot. The results show that the model is overall consistent and stable over different scenarios.

\begin{table}[htp]
    \centering
        \centering
        \begin{tabular}{l l l l l l}
        \toprule
              & & R2   & MSE  & MAE & MAPE\\
        \midrule
        OAS & March 2023 & 0.79 & 30.58 & 16.18 & 13.09\\
        & July 2022  & 0.79 & 25.94 & 14.98 & 2.07\\
        & March 2021 & 0.71 & 32.53 & 16.62 & 3.60\\
        & March 2020 & 0.68 & 28.61 & 15.66 & 1.97\\
        & June 2019  & 0.84 & 20.06 & 10.95 & 108.19\\
        \midrule
        Yield & March 2023 & 0.88 & 0.30 & 0.18 & 0.05\\
        & July 2022  & 0.88 & 0.35 & 0.22 & 0.07\\
        & March 2021 & 0.80 & 0.40 & 0.23 & 0.29\\
        & March 2020 & 0.72 & 0.38 & 0.19 & 0.14\\
        & June 2019  & 0.89 & 0.26 & 0.15 & 0.07\\
        \bottomrule
        \end{tabular}    
    \caption{Performance of CatBoost on the test data for both OAS and Yield as the targets for various snapshots.}
    \label{tab:test_evalmetric}
    \vspace{-6mm}
\end{table}

\subsection{Feature Importance using SHAP}
\begin{figure}
    \begin{subfigure}{\linewidth}
        \includegraphics[width=0.9\linewidth,height=6cm,trim={0 0 0 0.9cm},clip]{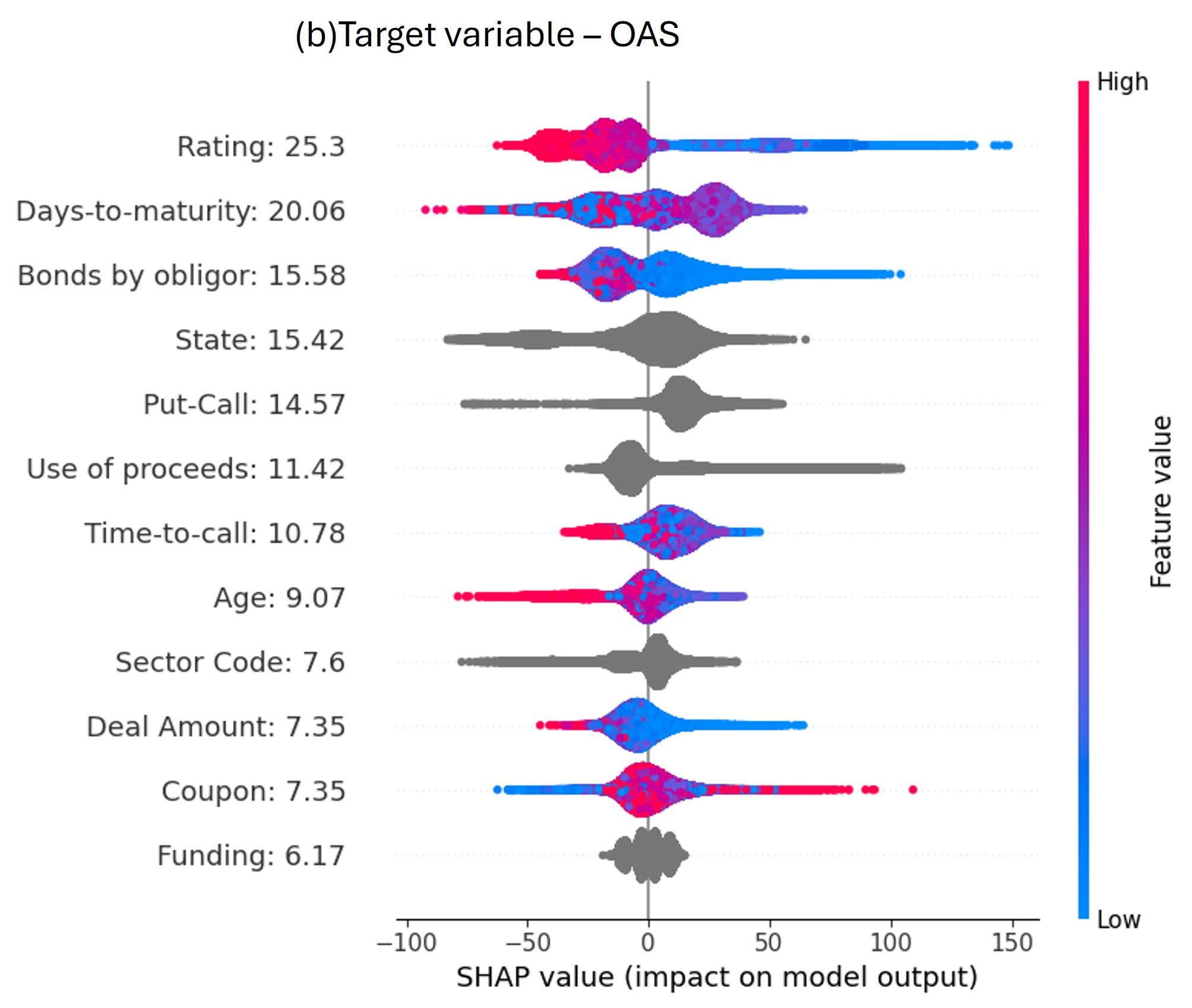}
        \caption{Values for target OAS}
    \end{subfigure}
    
    \begin{subfigure}{\linewidth}
        \includegraphics[width=\linewidth,height=6cm,trim={0 0 0 0.9cm},clip]{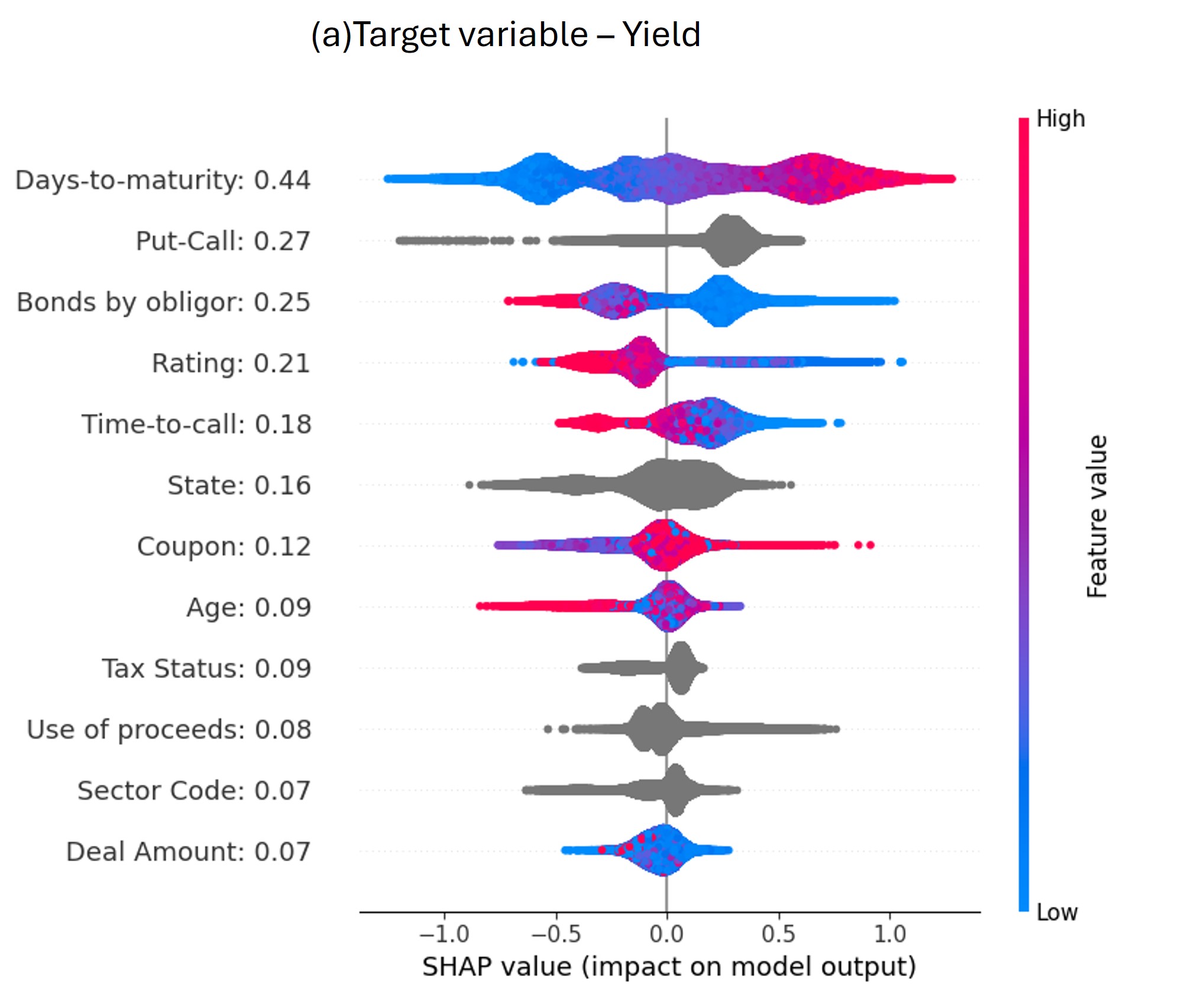}
        \caption{Values for target Yield}
    \end{subfigure}
    \caption{SHAPLY Value-feature importance in the CatBoost Model with Figure (a) Target variable-OAS and Figure (b) Target variable-Yield}
\label{fig:GBM_Shap}
\vspace{-6mm}
\end{figure}

Figure \ref{fig:GBM_Shap} shows the average (over the complete dataset) SHAP values as the feature importance for each variable. The Shaply values of the features impacting the target prediction are displayed for each of the target values - OAS and Yield separately. On average, ratings, days-to-maturity, and obligors are important features to predict the OAS, and days-to-maturity, put-call, and obligors are important features to predict the yield of the individual bonds. 

\subsection{Back-test Result}
\begin{figure*}[htbp]
    \centering
    \begin{subfigure}[b]{0.49\textwidth}
        \centering
        \includegraphics[width=\linewidth]{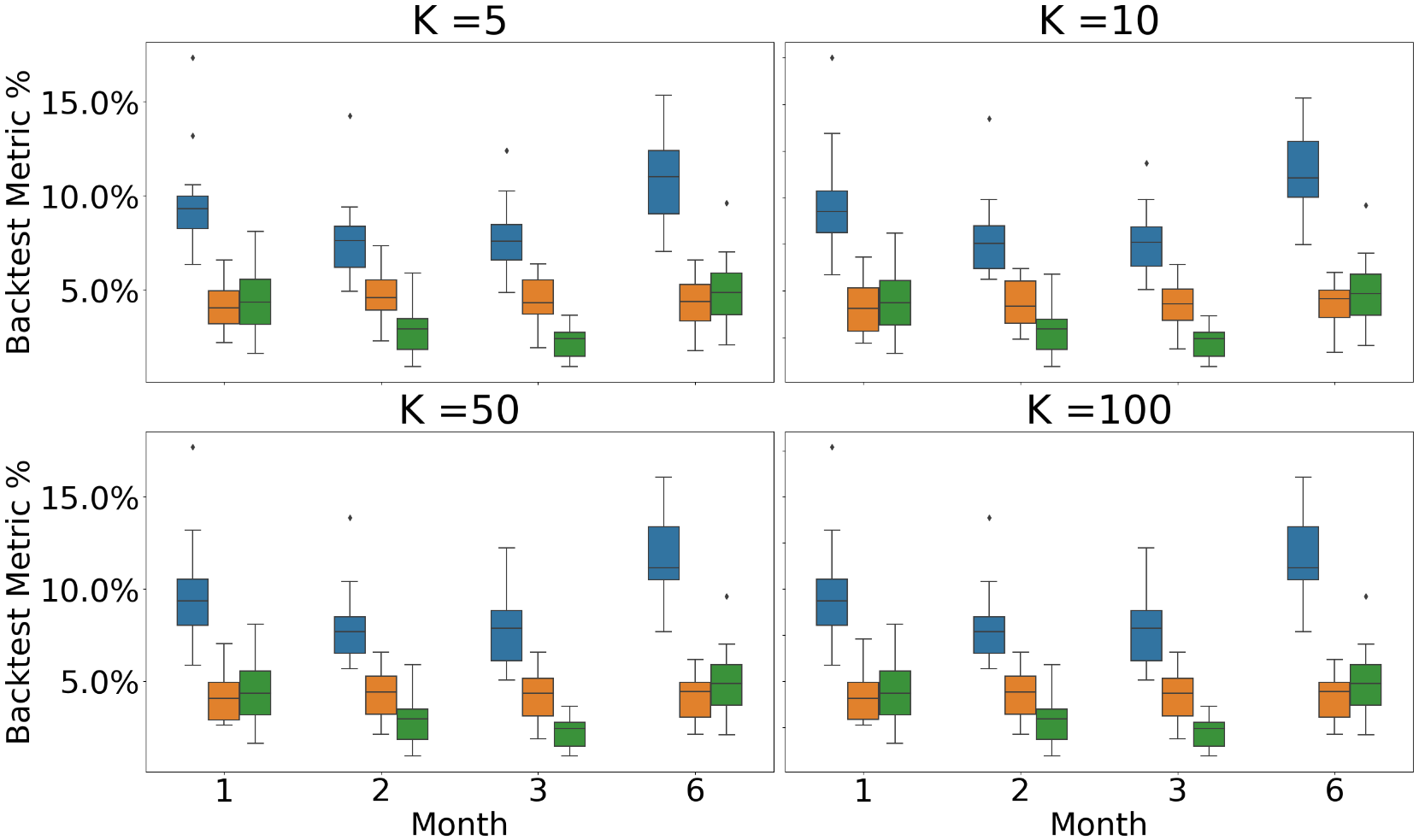}
        \caption{2024 Back-test Result}
        \label{fig:2024test}
    \end{subfigure}
    \hfill
    \begin{subfigure}[b]{0.49\textwidth}
        \centering
        \includegraphics[width=\linewidth]{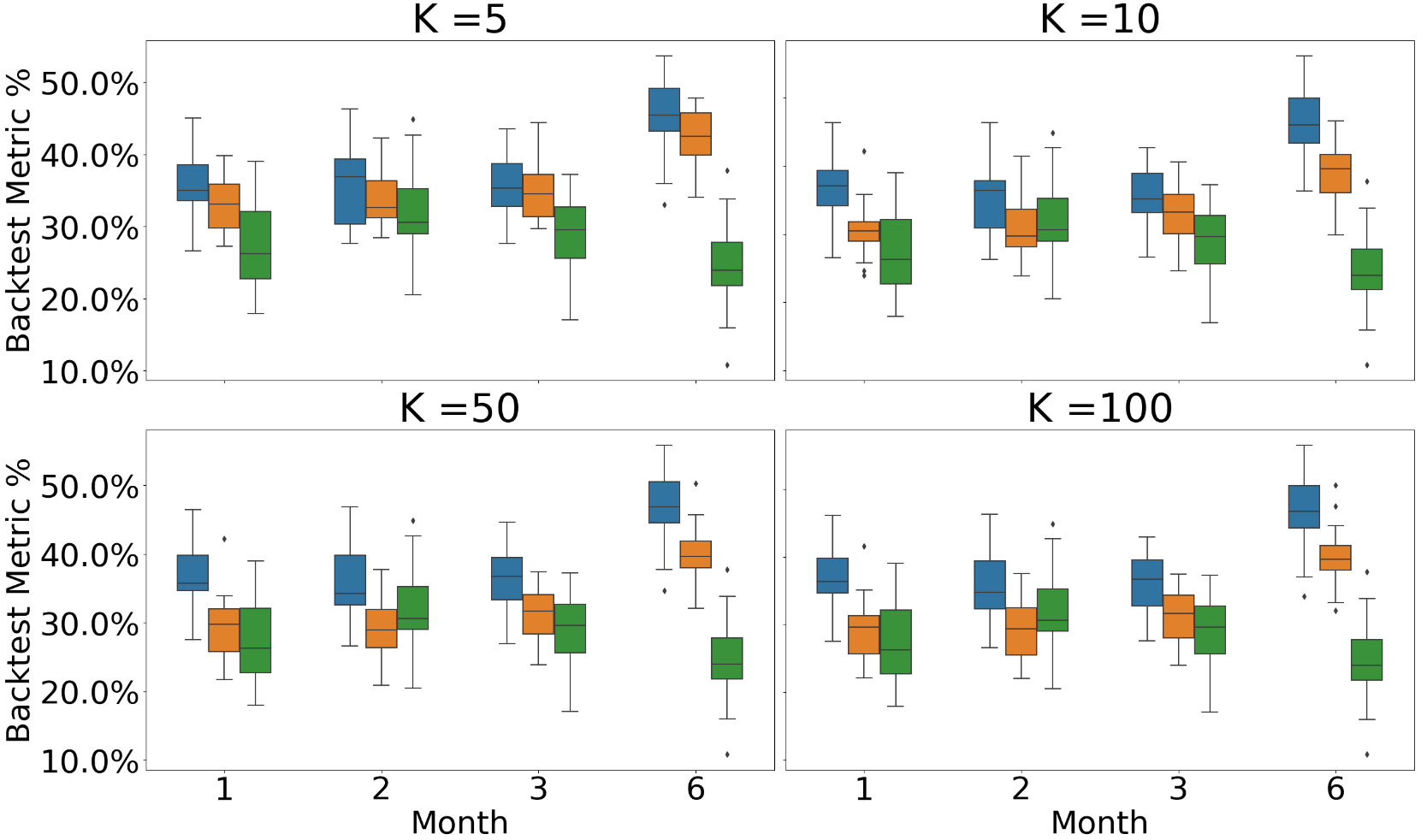}
        \caption{2023 Back-test Result}
        \label{fig:2023test}
    \end{subfigure}
    
    \begin{subfigure}[b]{0.49\textwidth}
        \centering
        \includegraphics[width=\linewidth]{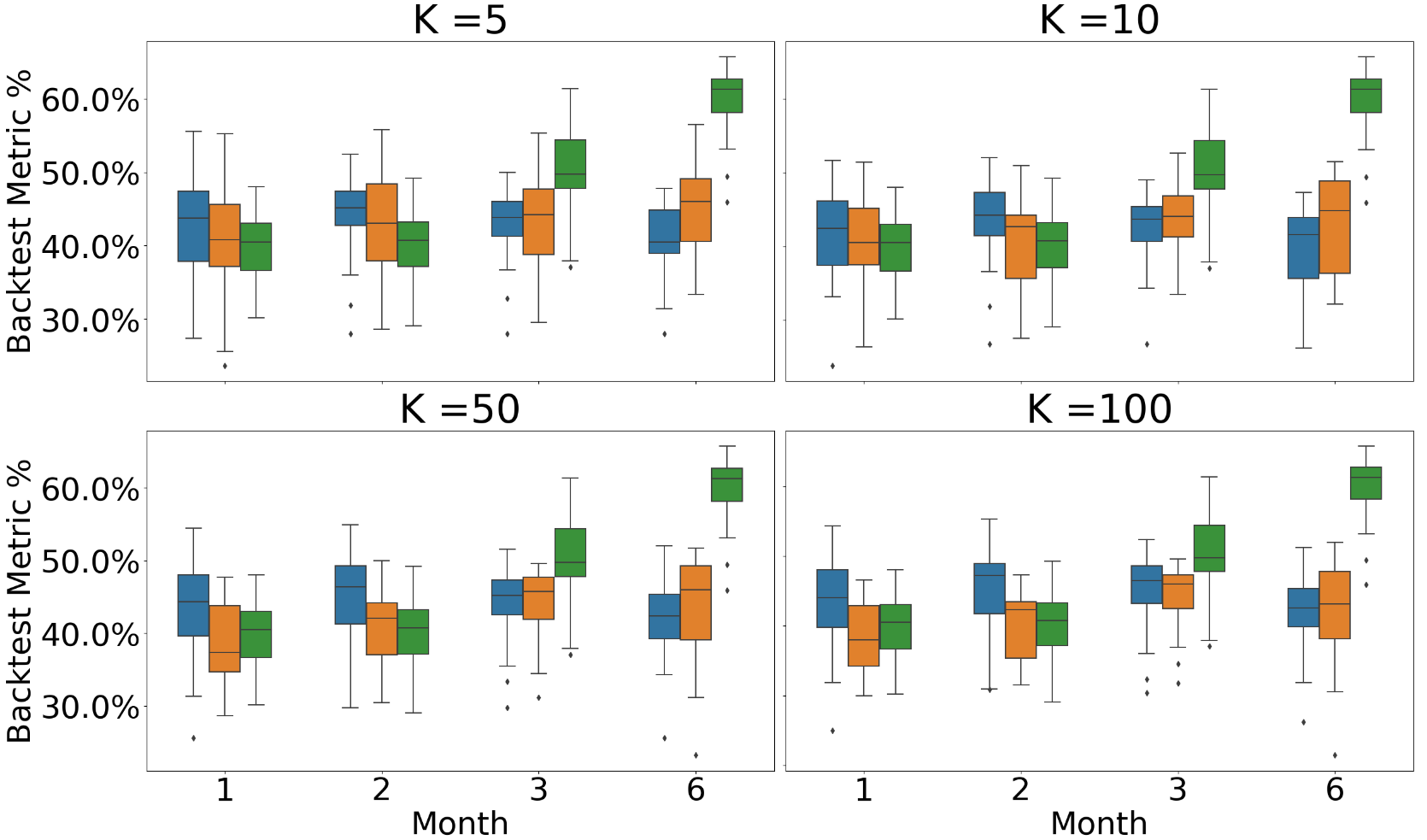}
        \caption{2022 Back-test Result}
        \label{fig:2022test}
    \end{subfigure}
    \hfill
    \begin{subfigure}[b]{0.49\textwidth}
        \centering
        \includegraphics[width=\linewidth]{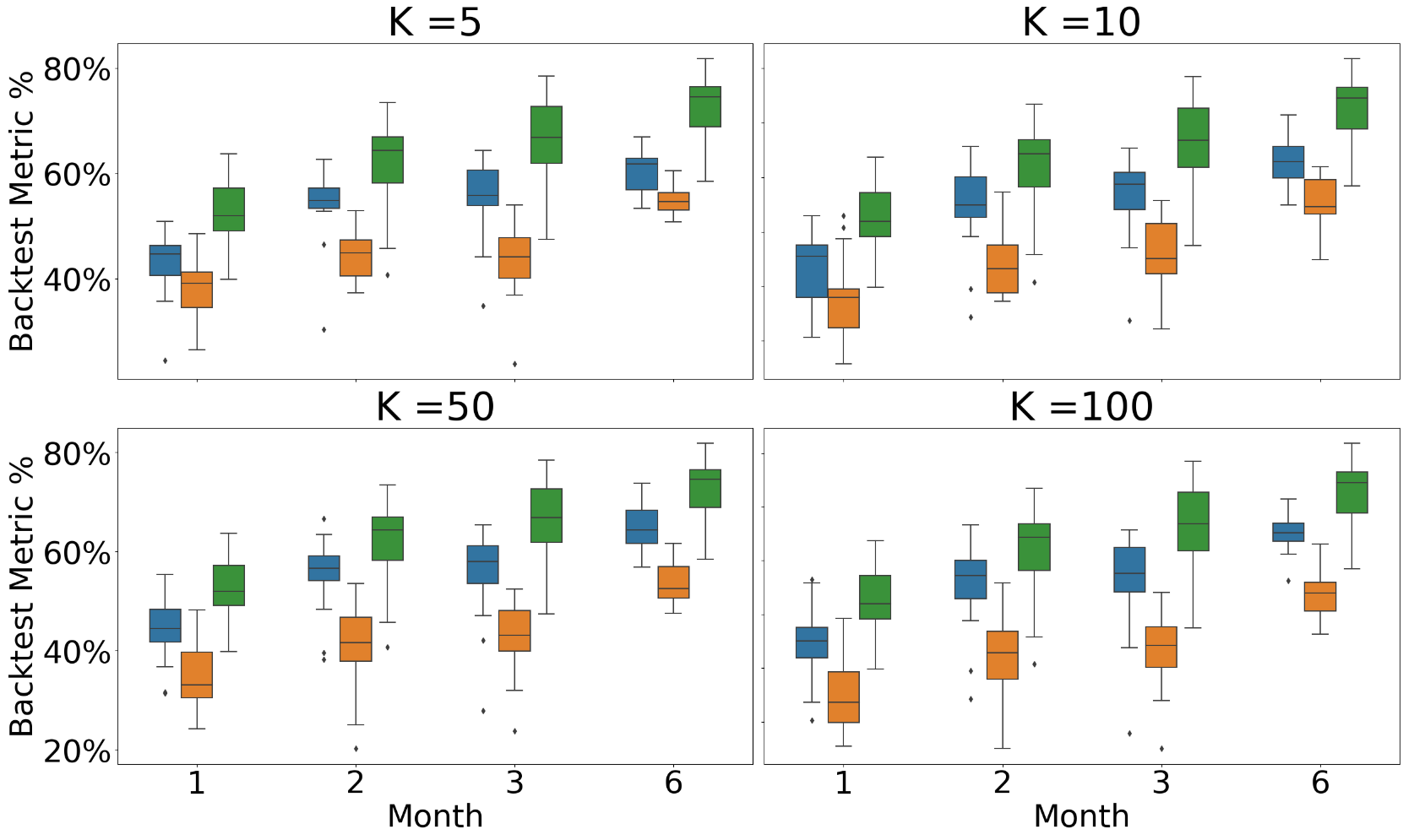}
        \caption{2021 Back-test Result}
        \label{fig:2021test}
    \end{subfigure}
    
    \begin{subfigure}[b]{0.49\textwidth}
        \centering
        \includegraphics[width=\linewidth]{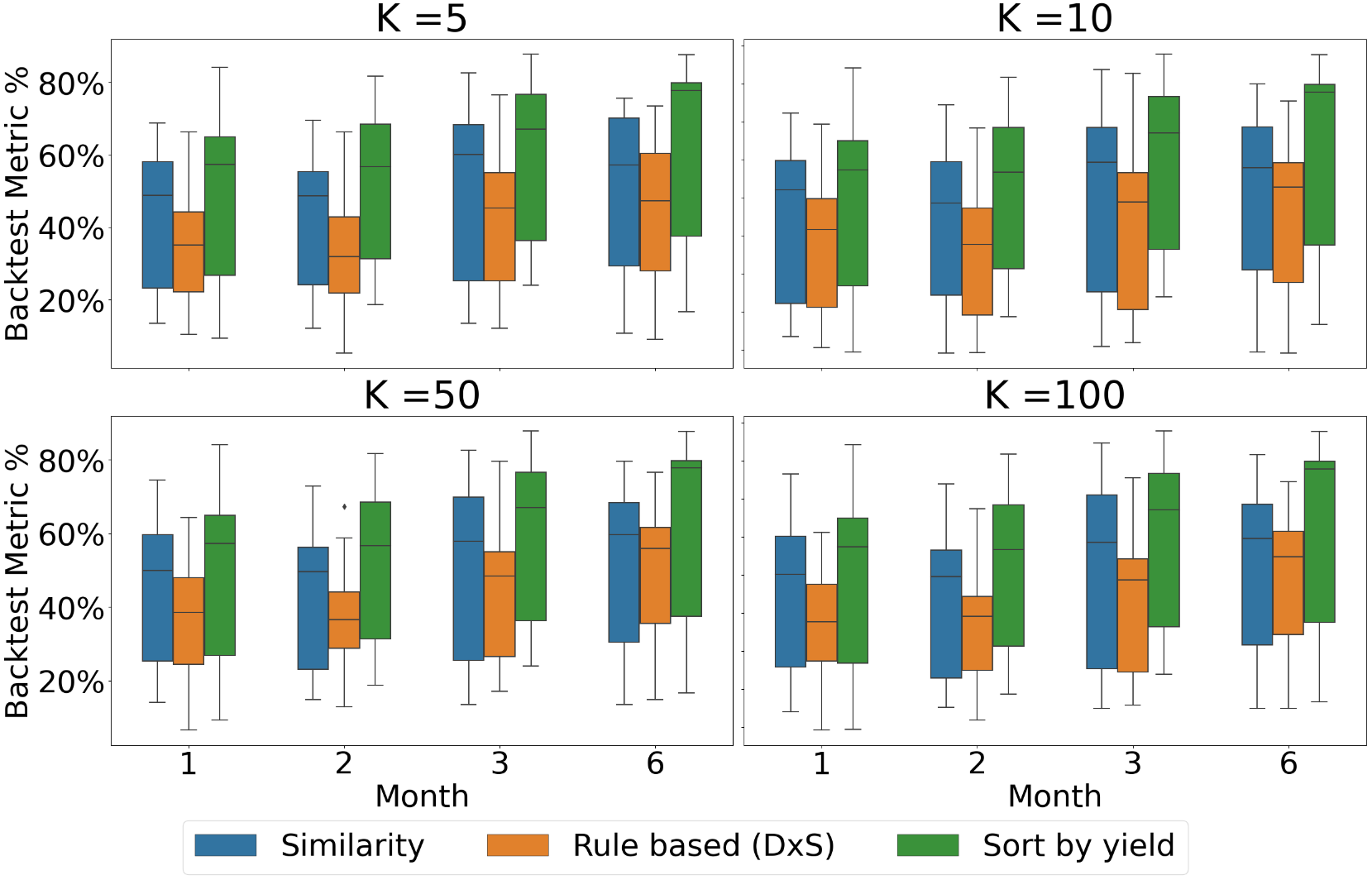}
        \caption{2020 Back-test Result}
        \label{fig:2020test}
    \end{subfigure}
    \hfill
    \begin{subfigure}[b]{0.49\textwidth}
        \centering
        \includegraphics[width=\linewidth]{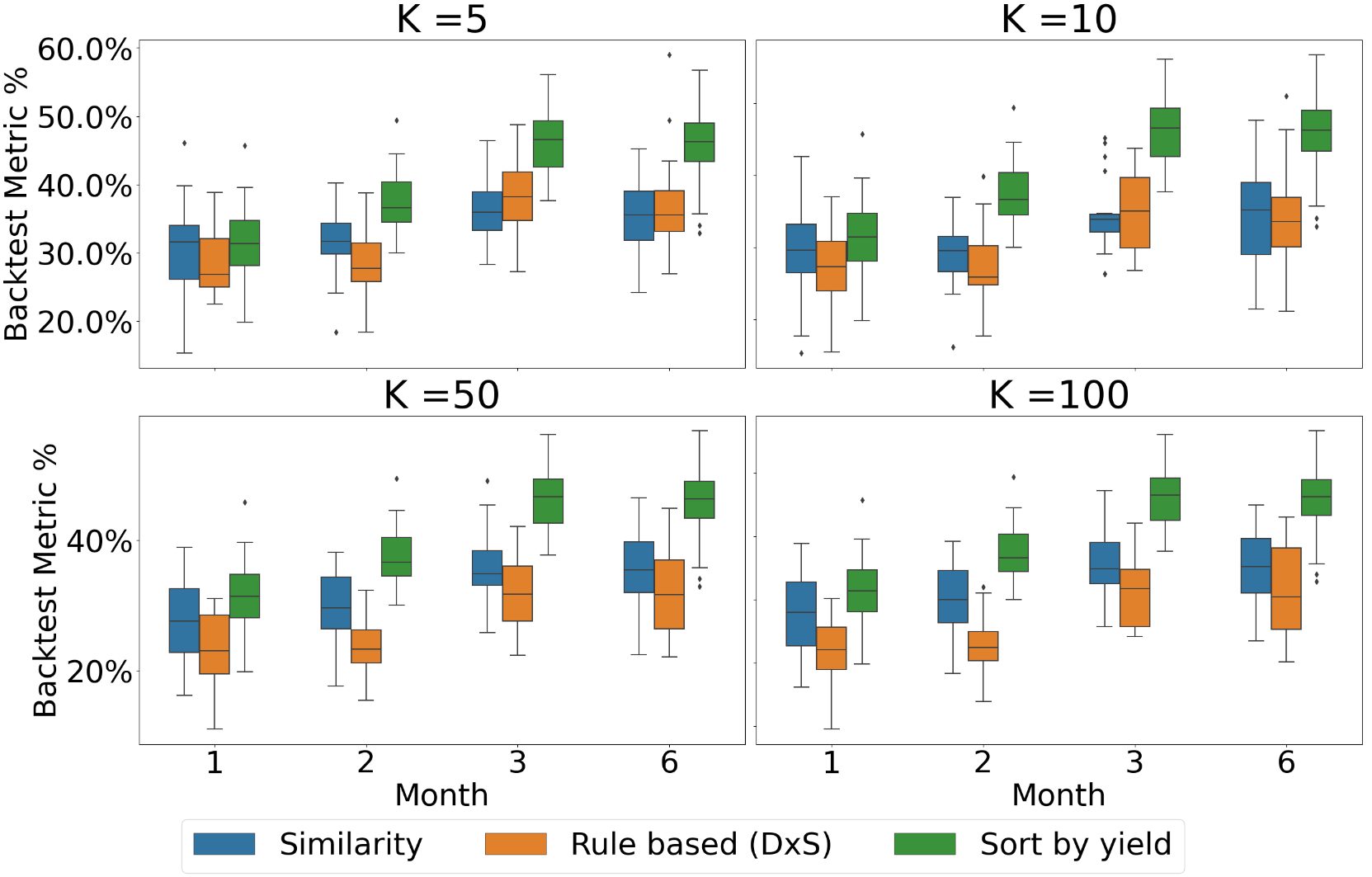}
        \caption{2019 Back-test Result}
        \label{fig:2019test}
    \end{subfigure}
    
    \caption{Back-test Results from 2019 to 2024. \textit{K is the number of nearest-neighbors (5,10,50 and 100) used to determine ranking for similarity and rule-based cohorts.}}
    \label{fig:alltests}
\end{figure*}

Figures (\ref{fig:2024test}-\ref{fig:2019test}) provide results from our experiments for back-testing for various scenarios.

For the November 2023-February 2024 scenario (Figure \ref{fig:2024test}), the muni bond market underwent a transition from volatility to recovery with yields peaking in October before declining as the Fed paused its rate hikes, and the March 2023 - September 2023 scenario (Figure \ref{fig:2023test}) captures the recent muni market rally in late 2023, characterized by declining yields. In these scenarios, overall similarity-based ranking outperforms other approaches.

For July 2022 - January 2023 scenario  (Figure \ref{fig:2022test}) covers a sharp increase in muni yields in 2022 due to rising US Treasury rates, which introduced volatility. During the rising rate and market booming environment, people would prefer more risky assets, which make higher-yielding bonds perform slightly better. The similarity ranking methodology shows the highest median from month 1 to month 2, reflecting strong performance, whereas sorting by yield and rule-based ranking methods show higher medians starting in month 3 with a broader interquartile range (IQR). This result shows that ranking methods have inconsistent performance in a period of rising yield indicating that all strategies are subject to significant fluctuations in a volatile yield environment.

The March 2021 - September 2021 scenario  (Figure \ref{fig:2021test}) captures the end of the muni market rally from 2019 to 2021, characterized by a low-interest rate environment and lower yields, where sorting by yield and similarity model-based ranking method show comparable medians while similarity-based ranking has relatively lower variability, suggesting both strategies were effective during the stable low-yield environment, and similarity has a more stable performance. 

During the March 2020 to September 2020 scenario  (Figure \ref{fig:2020test}), March 2020 marks the onset of the COVID-19 pandemic, leading to significant market volatility and a sharp drop in yields as the Federal Reserve cut interest rates to support the economy. Both similarity-based and sorting by yield rankings show higher medians and wider IQRs, suggesting that during the highly volatile market conditions at the onset of the pandemic, similarity-based strategies were effective.

Finally, for the June 2019 - December 2019 scenario  (Figure \ref{fig:2019test}), June 2019 was characterized by a continued low-interest rate environment, with the Federal Reserve pausing its rate hikes, leading to a stable or slightly declining yield environment in the muni market. The similarity-based ranking method shows a higher median compared to the rule-based method, while sorting by yield has stronger performance starting month 2 but displays high variability, indicating yield-focused strategies were effective in capturing returns but inconsistent.

Overall, the muni market's performance relative to Treasuries, along with interest rate trends and broader economic factors, have some impacts on the effectiveness of these strategies. In most cases, however, ranking based on our similarity model is better than the other two approaches, this means that the top rankers by the similarity approach also produce the highest relative return for its cohort in most generic bond trade orders. In cases, where sorting by yield has comparable or better performance, we discount for risk taken by selecting high-yielding bonds. 

\section{Conclusion}\label{sec:conclusion}
Given the illiquid nature of muni bonds identifying correct cohort for a given bond is crucial for different tasks such as pricing the bond and performing relative value analysis. In our work, we formulate the problem of order matching as a similarity learning problem where we first train a supervised similarity model using multi-output CatBoost regression with OAS and yield as the target variables. Using the trained model we extract proximity to identify neighbors for a given bond. We then couple this similarity with a relative value analysis approach where we compare the distance of a bond to its cohort based on our similarity model. 

We benchmark the similarity-aware relative value performance against a rule-based and a heuristic-based approach. Our proposed solution of using similarity-aware relative value outperforms the heuristic-based approach in different market scenarios over five years. Our approach also outperforms the simple rule-based method in some scenarios. We argue that buying the cheapest bond based on its yield can lead to the accumulation of excess risk. A similarity-based approach can account for this as the original model learns to identify bonds that are similar to each other based on their characteristics and target variables yield and OAS. In the future, we plan to enhance the concept of a generic group by advancing beyond basic filtering techniques and instead use concepts of prototypes and identify dynamic groups based using the similarity measure. 

\section{Acknowledgement}
The views expressed here are those of the authors alone and not of BlackRock, Inc. We thank Alim Abudu, Andrew Ang, Pratik Aswani, Maggie Blair, Gregory Bennett,  Sebastian Frank, Lisa Goldberg, Patrick Gorman, Bailey Hammer, Patrick Haskell, Mita Mehta, Alessio Muscara, Stefano Pasquali, Curtis Triece and Chad Wildman for their invaluable input and feedback that enhanced the quality of this work.

\bibliographystyle{unsrt}

\bibliography{sample-base}

\end{document}